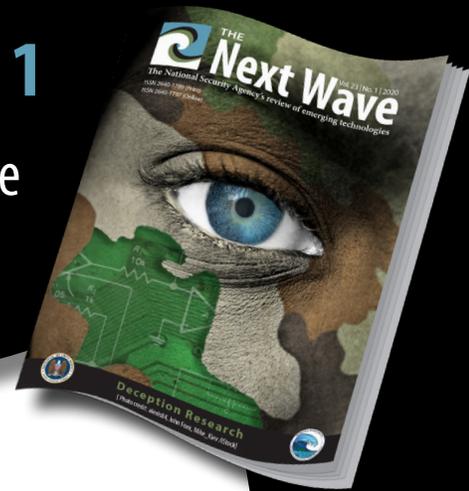

# DECEPTION AND THE STRATEGY OF INFLUENCE

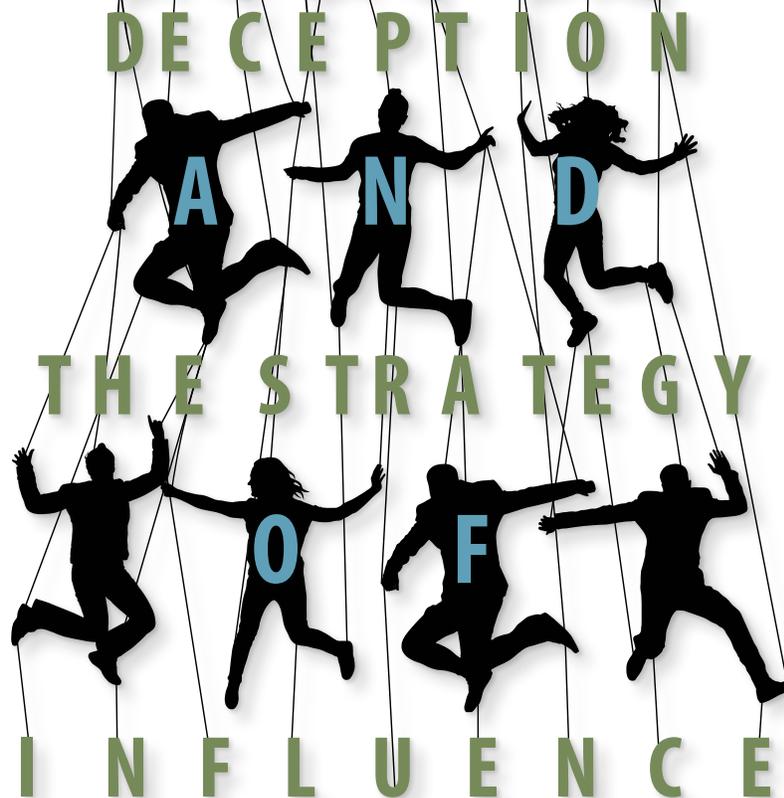

[Photo credit: iStock.com/Nosyrevy]

By Brian B., William Fleshman, Kevin H., Ryan Kaliszewski, Shawn R.

Organizations have long used deception as a means to exert influence in pursuit of their agendas. In particular, information operations such as propaganda distribution, support of antigovernment protest, and revelation of politically and socially damaging secrets were abundant during World War II and the Cold War. A key component of each of these efforts is deceiving the targets by obscuring intent and identity. Information from a trusted source is more influential than information from an adversary and therefore more likely to sway opinions.

The ubiquitous adoption of social media, characterized by user-generated and peer-disseminated content, has notably increased the frequency, scale, and efficacy of influence operations worldwide. In this article, we explore how methods of deception including audience building, media hijacking, and community subversion inform the techniques and tradecraft of today's influence operators. We then discuss how a properly equipped and informed public can diagnose and counter malign influence operations.

## History and background

The use of influence and deception as weapons is not a new concept. The famous general and philosopher Sun Tzu (545 BC–470 BC) said that "All warfare is based on deception" and "The supreme art of war is to subdue the enemy without fighting." Using information, both true and false, to confuse, divide, and demoralize opponents is a tactic that has been exploited for millennia.

### *The British agenda in Nazi Germany*

Between 1941 and 1943, Der Chef operated as the spokesperson of an illegal radio station in Nazi Germany, called GS-1 [1]. Der Chef acted as a loyalist to the Nazi cause and lambasted Nazi Party officials who he accused of being lazy, corrupt, and engaging in various sexual improprieties; meanwhile, he praised the bravery and devotion of German troops on the front line. In reality, Der Chef was a German refugee living in and recording and broadcasting from England. GS-1 was part of England's black propaganda engine, run by Sefton Delmer, which broadcast US jazz, German dance music, and sports scores, as well as reporting news to the public with a secret British agenda.

Der Chef would use reported local news and facts whenever possible to undermine the German populace's faith in Nazi leadership. Since facts are difficult to dispute, information used in this way is powerful and persuasive. Furthermore, by dispersing propaganda among music and news reports, Der Chef attracted new listeners and obfuscated his true intentions from his audience. Delmer described this approach to propaganda as "Cover, cover, dirt, cover, cover" while we refer to it as *pump-and-pivot*. Influence operators use this technique by drawing followers in through benign, popular content and then pivoting to malign influence.

### *The Communist agenda in Latin America*

In the 1960's, anti-American sentiment in Latin America led to footholds for communist elements. Compounding these problems, a letter that was signed by J. Edgar Hoover congratulating Thomas Brady for his efforts in the joint FBI/CIA operation to overthrow the Brazilian government was leaked to the press [2].

It turned out that the letter was fake, forged by the Czechoslovak Intelligence Service (CIS) to undermine US interests. The sensationalism of the story encouraged the media to release the story with little scrutiny or fact-checking. In addition, the anti-US sentiment of the population and confirmation bias caused the story to be met with little skepticism from the Latin American public. Predisposition and sensationalism make populations vulnerable to influence operations.

### *The Islamic State agenda in Libya*

On November 18th, 2014, CNN reported, "Fighters loyal to the Islamic State in Iraq and Syria are now in complete control of the city of Derna…The fighters are taking control of political chaos to rapidly expand their presence westward and along the coast, Libyan sources tell CNN" [3]. At that time, the caliphate seemed to be growing at an unprecedented rate. The Islamic State's strong expansion into Libya seemed to signal a groundswell of support and unity in the movement. Furthermore, this story from CNN was picked up by Time Magazine, the BBC, and Der Spegiel. The only problem—Derna was a hotbed contested by no less than three militant groups, and while the Islamic State had a presence there, their control was anything but complete. By controlling all of the information in the area, the Islamic State could write their own narrative of events and use mainstream media to disseminate it.

## Perceived legitimacy

People naturally associate with others with similar ideologies. These groups are often described as *echo-chambers or filter-bubbles*, amplifying the ideas common to the group while squashing the flow of discourse contrary to their shared beliefs. Once an influence operator has established a *persona*, or false presence, the insular nature of these bubbles stifles dissent and makes the group more susceptible to influence. In social media, these groups are referred to as *communities*; to an influence operator, they are an audience.

To conduct an influence operation, operators such as Der Chef need a receptive audience. The two ways to gain an audience are to build one, such as what Der Chef did by playing music, or to hijack an existing audience, as in the CIS and CNN cases. In both of these scenarios, the influence operator needs to appear as though they are providing a legitimate service to their targets.

In order to appear legitimate, it is important for the influence operator to avoid scrutiny. This is why the pump-and-pivot tactics are so common. During the *pump* phase, the operator manipulates the environment to increase legitimacy by, for example, appealing to a biased target, being the only source of information, using facts interspersed with falsehoods or directing targets to other compromised sources.

Deceptive media has been used to build legitimacy for some time, but the cost of producing quality material has traditionally reduced its scale. The impact on perceived legitimacy due to ubiquitous access to targeted synthetic media (e.g., deepfakes) generation will likely be profound. Malign actors will no longer need to draw on organic material that moves their narrative forward among their devoted following. They will instead be able to support their activities and narratives with synthetic content that appears to be factual evidence. This will decrease the time needed to both build their legitimacy and reach their malign influence goals. Successful operations will likely only be detected and mitigated by social media platforms as users will not have sufficient information to make an accurate assessment.

## Audience building

Social media has scaled audience building by providing targeted advertising, automation, and access to millions of users. These tools can be leveraged to precisely target demographics and build an audience out of previously disjoint subgroups [4]. Operators draw users to compromised information sources by providing information of interest with the intent of making the operator's persona and the information sources part of their targets' daily routines. During this time, the information sources provide a legitimate service (i.e., desired content). For example, recent reporting on Russia's Internet Research Agency (IRA) in 2017 demonstrated persistent attempts by IRA personas to cover local interest stories first, amplifying the spread of the stories with the help of automated accounts or *bots* [4]. By reporting first, and through the careful use of keywords, the bots landed at the top of trending news feeds and search results, building their audience.

Social media platforms facilitate the *pivot* phase by allowing users to reinvent their accounts without notifying the users who are within their network. From the point of view of the users it appears that a totally different actor has begun to contribute to their trusted information stream. This allows the operator to inject information from the compromised sources and amplify content arising from the community. Subsequently the community will move by itself, with the influence operator keeping the focus on the desired narrative. These behaviors of account reinvention can be observed in real time, but can be extremely difficult for the average user to observe in retrospect.

Synthetic media can play a substantial role in audience building—conversational bots can be leveraged to disseminate useful information at scale while engaging their audience, content sought by key audiences can be generated reducing cost and likelihood of being attributed, coordinated automated but realistically human bots can give the appearance of social consensus. The challenge of identifying these behaviors at scale to mitigate their impact or remove the associated campaigns entirely will be a persistent challenge for platforms for the foreseeable future.

## Media hijacking

In today's media environment, the rewards for possessing timely, exclusive reporting on a topic can incentivize publication before rigorous fact-checking is available. This is particularly true of content that is emotionally charged—sensationalism drives increased readership and engagement. For example, the increased risk to journalists in militant sites reduces the availability of professional journalism in a region, but battlefield reporting is valuable news. By providing professional quality reporting in such a region, influence operators can have their reports repeated and amplified by the international press, producing an immediate audience.

The rush-to-publish environment facilitates influence operators use of synthetic media to amplify deceptive narratives. It is now possible to generate realistic video and audio of a well-known personalities at minimal cost. This capability will likely be leveraged by influence operators to divert attention from legitimate but damaging news stories as well as create confusion in times of uncertainty. Media companies will need to balance the desire to be first-to-publish with the possibility of providing a platform to influence operators.

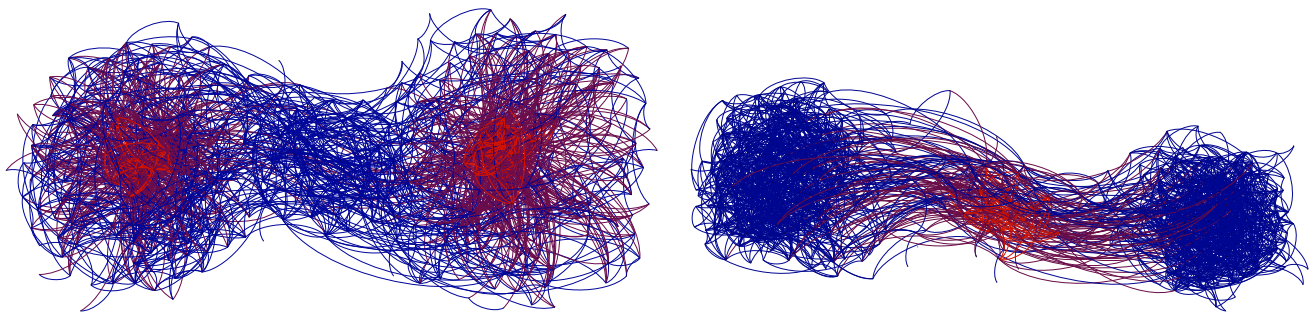

**FIGURE 1.** This simulated data shows the flow of information from influence operators into a two-sided discourse. Vertices represent users, red edges represent communication from embedded operators, blue edges represent communications from legitimate users. (Left) The embedded personas are attempting to influence both sides of the discourse from the community cores. (Right) The embedded personas act as bridges between the communities in order to develop malign confrontation.

## Community subversion

Some influence operations are not as direct as Der Chef decrying the Nazi leadership or the Islamic State controlling the narrative of battle around Derna. Influence operators can use the perceived presence of numbers to change the narrative of a community, a tactic which is referred to as *community subversion*. For example, the Saudi Arabian government was accused of using bots to undermine anti-Saudi hashtags and inflate pro-Saudi positions surrounding conflict with Qatar [5], and Iran has been accused of using more than 140 Reddit accounts to promote anti-Saudi, anti-Israeli, and pro-Palestinian narratives [6].

Those two examples of community subversion illustrate *bolstering* and *degrading* of communities. Bolstering a community is a subversion technique where influence operators artificially increase support in order to embolden legitimate users. Degrading a community is a tactic where operators sow division within the community.

Influence operators can also interfere with a community through a *denial of service attack*. By flooding the community with noise, they can either trigger a platform's automated spam filter or prevent legitimate users from communicating in an organic way. This was seen firsthand in 2014 when bots entered a human rights Twitter community centered on protests in Mexico and filled it with spam [7], preventing protesters from coordinating to avoid police. The left-hand plot of figure 1 graphically shows how influence operators would be situated in the above attacks.

Other examples of community subversion come from Russian IRA influencers, who in 2016 were observed operating on both sides of the Black Lives Matter hashtag [8]. By doing so, the influencers directly inflamed the discourse on both sides by moving both conversations to the extremes.

Then, in 2018, researchers observed Russian IRA influencers acting as a bridge between polarized groups in the vaccination debate [9]. They were forcing communication and specifically argumentation between groups of people on opposite sides of the issue by using the hashtag #VaccinateUS for both pro-vaccine and anti-vaccine content. Furthermore, the #VaccinateUS tweets generally included other emotionally charged topics from US culture in order to maximize division. The right-hand side of figure 1 shows an example of such behaviors—connecting two communities that would otherwise be loosely connected or disconnected.

Automated community-to-community interactions can now be scheduled and convincingly generated based on current conversation. Present and timely synthetic media will increase the chance that these bridging operations succeed. Community members should question new narratives entering their networks from previously unknown accounts. This is particularly true for controversial or confirmation bias affirming narratives.

## Diagnosis of influence operations

Efforts to detect influence operations leverage the behaviors that result from operators' desire to minimize their fingerprints on the larger conversations. The goals of minimizing direct involvement and trying to build an audience are often in contention over the course of an influence campaign. This results in opportunities to identify coordinated networks of accounts. The basic tools for these observations are *community detection* and *content analysis*.

Community detection is an algorithmic way to detect internal structure in a network graph such as comment, email, or retweet graphs. In particular, if one constructs their graph in such a way as to indicate positive sentiment between nodes, then such a graph can be viewed as an indicator for likely confirmation bias. Content analysis, such as topic modeling [10, 11] or text summarization algorithms, can isolate the themes in the discourse and be used to understand the narrative and focus within a community as well as the flow of discussion between communities.

While the boundary between social media and mainstream reporting is becoming ever more porous, efforts to mitigate the spread of influence operations should pay close attention to the beginnings of discourse. Integration of dynamic content analysis can highlight the construction of new narratives, and particular attention should be focused on narratives with extreme amplification during this time period. This sort of analysis is particularly important regarding stories around which the information environment is particularly constrained.

As a discourse matures, accounts with deceptive behavior should be analyzed closely. Aberrant behavior such as removing a great deal of previous content, changing outward appearance, or a distinct change in quality or focus of shared content can indicate a pump-and-pivot. Established communities that suddenly shift focus or trigger flags such as spam filters can be indications of a pump-and-pivot or community subversion. The application of social bot classifiers can help separate artificial amplification from organic growth, highlighting accounts that attempt to inject themselves into the discourse.

*Brigading*, or accounts that join a community for purposes other than joining the discourse, could signal a community subversion effort, especially if the brigading is coordinated or in excessively large numbers. More specifically, if the number of interactions between two polarized communities increases, then further analysis can be done to investigate whether the increase is natural or caused by a deceptive force.

## Detecting influence operations through technology

Another way to identify influence operations is by how they choose to interact with social media. Users interact with a social media platform via a *client*. While many users operate with first party clients, a number of third-party clients exist to facilitate automation, provide a different look and feel, allow for management of multiple accounts across different platforms, and display analytics of audience engagement [12].

Raw access to the platform's application programming interface (API) can provide the ability to spoof geolocations, IP addresses, and timing of posts to appear to be elsewhere in the world [13]. Influence operators frequently establish personas in different countries to conduct influence operations more effectively [4] and build legitimacy. Operators can create a custom client to increase their efficiency, allowing one person to control dozens or hundreds of accounts with varying degrees of automation.

We refer to the collection of clients used by an account to interact with a social media platform as its *technology stack*. Analysis of a technology stack for a specific account can help identify automated accounts via the presence of bot clients [14]. One hypothesis is that a human being in charge of an account will either be satisfied by standard, first-party or popular third-party, social media clients or will have some reason to seek out nonstandard clients. For each nonstandard third-party client there is likely a community of users that use that client.

Fingerprinting of technology stack communities can be done at scale by forming a bipartite graph of technologies and users within a social media platform. Detecting and removing outlier user and client nodes generates connected subcomponents of the graph. After dimensionality reduction, low-dimensional clustering algorithms can detect *clusters of clients* which are used by similar users and *clusters of users* with similar client usage. This allows for efficient analysis of subsets of users based upon their technology choices, such as in figure 2. The use of nonstandard clients with restrictive access requirements may indicate a relationship within the subset of users.

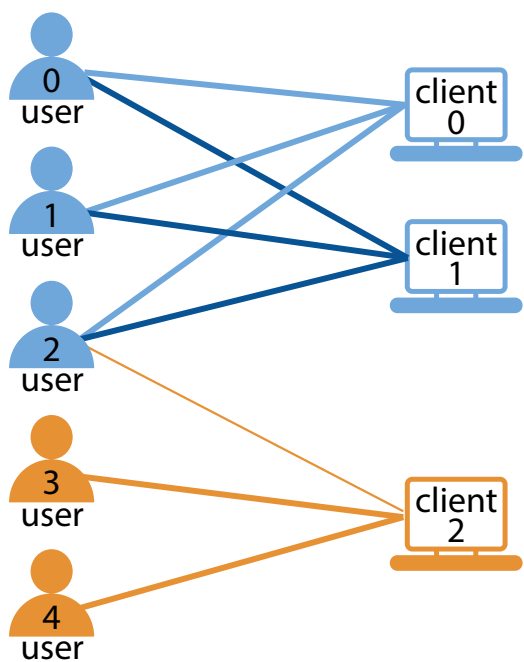

**FIGURE 2.** This bipartite graph illustrates user and client interactions with line thickness indicating the proportion of a user's total interactions with a social media platform using a particular client. Two clusters are shown, colored blue and orange respectively. Users 0, 1, and 2 use the blue technology stack (clients 0 and 1) with similar frequency, even though user 2 will rarely use client 2. Users 3 and 4 use the orange technology stack (client 2) exclusively and more frequently than user 2.

Identifying suspicious actors via their technology stack allows one to detect and mitigate influence operations early. This assists in preventing malign influence operators from achieving their goals and increases the cost necessary to deceive a target audience. Even if a campaign cannot be prevented using the above technique, post-hoc analysis can provide critical indicators which can be used by government or industry to increase the cost of future operations by preventing reuse of technology stacks by operators.

## Conclusion

Tools to create, curate, and automate convincingly human-created media (i.e., audio, text, image, and video) are readily available. These tools are already being used by influence operators to gain legitimacy, build their audiences, hijack traditional media, and subvert communities. This creates a persistent challenge for users, platforms, and media companies to address. A commitment on the part of the platforms to maintain technological solutions to identifying state-of-the-art synthetic media and influence campaigns, automate responses to identified activities, and provide context to the users would help mitigate these activities. Easy access to tools and knowledge to identify and respond to influence operations will help limit their impact. A collaboration between media companies and technology platforms to help identify synthetic media before broadcast will help reduce the likelihood of such broadcasts being leveraged by influence operators. In short, education, collaboration, and technology can be used together to help blunt the impact of synthetic media on public discourse.

Public discourse is, and has been, consistently influenced by malign actors. The growth of social media companies over the last decade has created a new dynamic in this system with which society has yet to find balance. The first tool necessary for finding that balance is knowledge—knowledge of when actors are trying to influence, knowledge of actors' intent in what to influence, and knowledge of who is acting in concert with an influence campaign. This knowledge cannot come from historical analysis alone as influence operations are dynamic from account creation to daily targeting. The flow of specific discussions through a network can be used to identify the target events and communities of an influence operation. Examining the specific behaviors of accounts can identify out-of-band coordination and automation. Together, these can give the public the knowledge to confidently interact with social media.

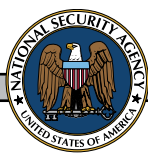
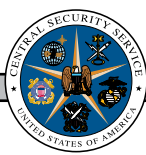
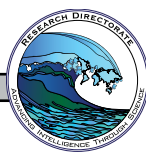